# Livelayer: A Semi-Automatic Software Program for Segmentation of Layers and Diabetic Macular Edema in Optical Coherence Tomography Images

Mansooreh Montazerin[1], Zahra Sajjadifar[1], Elias Khalili Pour[2], Hamid Riazi-Esfahani[2], Tahereh Mahmoudi[3], Hossein Rabbani[4], Hossein Movahedian[5], Alireza Dehghani[5], Mohammadreza Akhlaghi[5], Rahele Kafieh[4*]

[1]Department of Electrical and Computer Engineering, Isfahan University of Technology, Isfahan, Iran

[2]Retina Service, Farabi Eye Hospital, Tehran University of Medical Sciences, Tehran, Iran

[3]Department of Biomedical Systems and Medical Physics, Tehran University of Medical Sciences, Tehran, Iran

[4]Medical Image and Signal Processing Research Center, School of Advanced Technologies in Medicine, Isfahan University of Medical Sciences, Isfahan, Iran

[5]Department of Ophthalmology, Isfahan University of Medical Sciences, Isfahan, Iran

## Abstract

Given the capacity of Optical Coherence Tomography (OCT) imaging to display symptoms of a wide variety of eye diseases and neurological disorders, the need for OCT image segmentation and the corresponding data interpretation is latterly felt more than ever before. In this paper, we wish to address this need by designing a semi-automatic software program for applying reliable segmentation of 8 different macular layers as well as outlining retinal pathologies such as diabetic macular edema. The software accommodates a novel graph-based semi-automatic method, called "Livelayer" which is designed for straightforward segmentation of retinal layers and fluids. This method is chiefly based on Dijkstra's Shortest Path (SPF) algorithm and the Live-wire function together with some preprocessing operations on the to-be-segmented images. The software is indeed suitable for obtaining detailed segmentation of layers, exact localization of clear or unclear fluid objects and the ground truth, demanding far less endeavor in comparison to a common manual segmentation method. It is also valuable as a tool for calculating the irregularity index in deformed OCT images. The amount of time (seconds) that Livelayer required for segmentation of ILM, IPL-INL, OPL-ONL was much less than that for the manual segmentation, 5s for the ILM (minimum) and 15.57s for the OPL-ONL (maximum). The unsigned errors (pixels) between the semi-automatically labeled and gold standard data was on average 2.7, 1.9, 2.1 for ILM, IPL-INL, OPL-ONL, respectively. The Bland-Altman plots indicated perfect concordance between the Livelayer and the manual algorithm and that they could be used interchangeably. The repeatability error was around one pixel for the OPL-ONL and < 1 for the other two. The dice scores for comparing the two algorithms and for obtaining the repeatability on segmentation of fluid objects were at acceptable levels.

## Introduction

Optical Coherence Tomography (OCT) is a non-invasive, relatively inexpensive imaging technique which is based on low-coherence interferometry and captures high-resolution multi-dimensional images from biological tissue especially the retina [1]. Macular OCT images are widely used to assist ophthalmologists in diagnosing ocular deformities such as Diabetic Macular Edema (DME), Age-Related Macular Degeneration (AMD), glaucoma and retinal vascular accidents [2, 3]. In addition to that, their interesting applications in diagnosis and effective treatment of neurodegenerative diseases like Multiple Sclerosis (MS) and Neuromyelitis Optica (NMO) has attracted the neurologists [4].

Ocular OCT images provide cross-sectional data from inter-retinal layers which are distinguishable by contrasting their intensities. These layers typically lose their standard features (like texture, thickness and location) with the occurrence of different diseases and measuring the quantitative amount of their structural conversion, provides instructive information about the type, severity and the must-be-employed treatment procedure of that disease [5]. Furthermore, macular OCT is the standard test for detection of intraretinal and subretinal fluid. It is also an essential modality for evaluating the subsequent resolution of accumulated fluids as a response to treatment [6].

OCT data may be acquired in different anatomical locations in the eye. In the macular OCTs, a set of 2-dimensional (2D) cross-sectional B-Scans are acquired and stacked to generate a 3-dimensional (3D) macular cube. Besides, a 2D circumpapillary - retinal nerve fiber layer (cp-RNFL) data may be taken from the area surrounding the bundle of nerve fibers at the back of the eye, called optic nerve head (ONH) [2].

Macular OCT images carry a great deal of data which needs to be quantified to provide interpretable values. The main issue in this regard is segmentation of retinal layers and localization of abnormalities like intraretinal or subretinal fluid accumulations. For segmentation of inter-retinal layers and fluids in each B-Scan, assorted manual, semi-automatic and full-automatic approaches have been suggested [7-11]. The manual segmentation method is both time-consuming and exposed to probable observer errors and therefore, automatic and semi-automatic algorithms have been introduced to solve these problems. Full-automatic methods for segmentation of retinal OCT images are prone to unavoidable errors in the presence of less predictable matters such as noisy, low-quality images and in B-Scans of patients with marked macular or RNFL changes [12-15]. Accordingly, a semi-automatic approach is considered as an in-between method to overcome the disadvantages of manual and automatic methods and to take advantage of an expert's knowledge of a correct delineation. Table 1 summarizes a list of previous works on semi-automatic OCT segmentation. However, available semi-automatic algorithms suffer from noticeable constraints because they mostly support only one data format, are unable to suggest alternative segmentation methods to the users, lack an effective implementation of a detailed process on the input data like denoising or filtering and finally, are not integrated in an open-source software environment to serve facilitated segmentation of OCT images.

Table 1. Description of the preceding algorithms for OCT semi-automatic layer and fluid segmentation

| Algorithm's Name | Input | Number of Detected Layers | Location of Segmentation |
|---|---|---|---|
| EdgeSelect [16] (2013) | SD-OCT (Heidelberg Spectralis) | 3 Retinal Layers/4 Boundaries (ILM,IS/OS,RPE,BM)* | Macula |
| Kago-Eye2 [17] (2018) | SD-OCT (Heidelberg Spectralis) | 2 Borders (C-S,S-H) | Choroid |
| Zhao's Method [18] (2012) | SD-OCT (Heidelberg Spectralis) | 9 Retinal Layers (ILM, NFL/GCL, IPL/INL, INL/OPL, OPL/ONL, ELM, IS/OS, OS/RPE, RPE/CH)* | Macula |
| Liu's Method [19] (2018) | SD-OCT (Heidelberg Spectralis) | 8 Categories (ILM, NFL-IPL, INL, OPL, ONL-ISM, ISE, OSE-RPE, Fluids)* | Macula |
| SAMIRIX [20] (2019) | SD-OCT (Heidelberg Spectralis) | 9 Boundaries (ILM, RNFL-GCL, IPL-INL, INL-OPL, OPL-ONL, ELM, IS/OS, OPT-RPE, BM)* | Macula |

*Each retinal layer's abbreviation stands for as follows: ILM: Inner Limiting membrane, NFL: Nerve Fiber Layer, GCL: Ganglion Cell Layer, IPL: Inner Plexiform Layer, INL: Inner Nuclear Layer, OPL: Outer Plexiform Layer, ONL: Outer Nuclear Layer, ELM: External Limiting Membrane, IS/OS: Inner and Outer Segment, RPE: Retinal Pigment Epithelium, BM: Bruch's Membrane, CH: Choroid.

The purpose of our study is to design a piece of software to provide semi-automatic segmentation of layers and fluids in macular OCTs. Our proposed software considerably resolves the above-mentioned difficulties that other similar algorithms have faced. By performing an accurate and user-friendly segmentation, it helps clinicians to evaluate structural changes, especially at inner macular and cp-RNFL, with a more reliable data. It is also practical in supplying the gold standard data set needed for design and evaluation of full-automatic methods [21, 22].

To measure the validation of this software, we tested its performance on macular OCT B-Scans of eyes from healthy controls as well as those belonging to patients with DME diagnosis by calculating the agreement (to assess inter-rater variability and to decide whether the proposed technique could be substituted with the manual segmentation) and repeatability of the proposed method [23]. Moreover, to demonstrate one important application of the proposed method in clinical works, we designed a particular section for computation of the irregularity index in deformed OCT images, which could be used to assess irregularities frequently seen in many layered ocular structures (e.g. retina, iris, cornea) under pathologic circumstances such as DME, AMD, Epiretinal Membrane (ERM) [24], Vitreoretinal Interface abnormalities, Fuchs Heterochromic Iridocyclitis (FHI), Fuchs uveitis in Anterior Segment OCT images [25], and corneal dystrophies. Because the manual algorithms yield erroneous and unreliable segmentation of the retinal layers in macular OCT images, they could face major problems with respect to gauging this parameter and this is another situation where the semi-automatic and full-automatic segmentation algorithms are deemed extremely helpful.

In what follows, we explain about our main method and the overall structure of our designed software. Two disparate datasets (normal and DME) are introduced and the outcomes produced by applying our algorithm over the layers and fluid objects of these datasets are investigated by assessing the agreement and repeatability of the proposed method. Subsequently, the tables indicating validation results are provided in detail and an in-depth discussion of this research's outcomes, performance and efficiency in comparison to other heretofore suggested methods is summed up in the closing section.

## Materials and Methods

### Study population

To signify the clinical performance of the proposed method on retinal layers of a collected normal dataset and to calculate the agreement and repeatability of our algorithm's layer segmentation section, fifty macular OCT B-Scans from 16 normal eyes were selected in this study. So as to identify the clinical performance of this method on the fluid objects appearing in abnormal datasets, we enrolled twenty eyes from 19 patients with the diagnosis of diabetic macular edema (DME) [26]. All patients had clinical and OCT-based diagnosis of DME. OCT examinations were performed using Spectralis EDI-OCT and Spectralis SD-OCT for the normal and DME patients, respectively. Both datasets were gathered using Heidelberg Eye Explorer (HEYEX) version 5.1 (Heidelberg Engineering, Heidelberg, Germany) by a trained technician with an automatic real-time (ART) function for image averaging and an activated eye tracker in a room with normal light. For macular volumes, 61 horizontal B-scans (each with 512 A-Scans, with ART of 9 frames, axial resolution of 3.8 mm) with a scanning area of 6mm × 6mm focusing on the fovea were taken. The normal population's age ranged between 21 and 46 (Mean (±SD) = 30.81 ± 7.04) and 62.5 per cent of them were female. The DME group's age were between 57 and 84 (Mean(±SD)=68.11(±8.58)) and 74 per cent of them were female. The current study was carried out in accordance to the tenets of the declaration of Helsinki and a proof of concept was obtained from the patients.

**An overview of the proposed software**

Two primary goals of this study are elaborated in the next sections. We first explain an overview of the proposed software and the embedded semi-automatic segmentation algorithm and then, describe the agreement and repeatability of the studying process using the supporting dataset of OCTs from healthy controls and DME sufferers. Our proposed software consists of five independent tabs, each responsible for a specific function. First of all, in the "File" tab, the user could open the desired data format (i.e. .mat, .vol or .octbin) which is then converted to a .mat file in order to be compatible with this software and any other substituted MATLAB [27] code. The next three tabs are designed for layer and fluid segmentation of macular B-Scans. Starting with the "Manual Layer Segmentation" tab, a complete manual segmentation of retinal layers using the interactive freehand over each B-Scan is provided and the resulted curves that could be mainly used for construction of Gold Standards are produced. Meanwhile, the "Auto Layer Segmentation" tab is responsible for the software's main algorithm, a graph-based semi-automatic segmentation, termed "Livelayer". Two other options embedded in this tab are a layer correction procedure used for correction of the faulty boundaries, and a grid-based segmentation method which works by extrapolating the intended boundary from a limited number of clicks on each B-Scan. Fluids' identification and localization is the "Fluid Segmentation" tab's duty that adopts both manual (interactive freehand object) and semi-automatic (Livelayer) techniques. The last tab named "Peripapillary" is created for analysis of circumpapillary scans. Our suggested method removes the curvature, localizes the vessel shadows, eliminates the shadows, and finally provides Livelayer semi-automatic segmentation of the peripapillary boundaries. The extracted information from above-mentioned tabs is finally saved in a MATLAB ".mat" file to be readable in next applications. Fig. 1 represents our proposed software's main tabs. A brief illustration of conducted operations in our software for macular and circumpaillary images from accepting the input file with a .mat, .vol or .octbin format to outputting both the segmented B-Scan and the complementary information with respect to it is shown in Fig. 2.

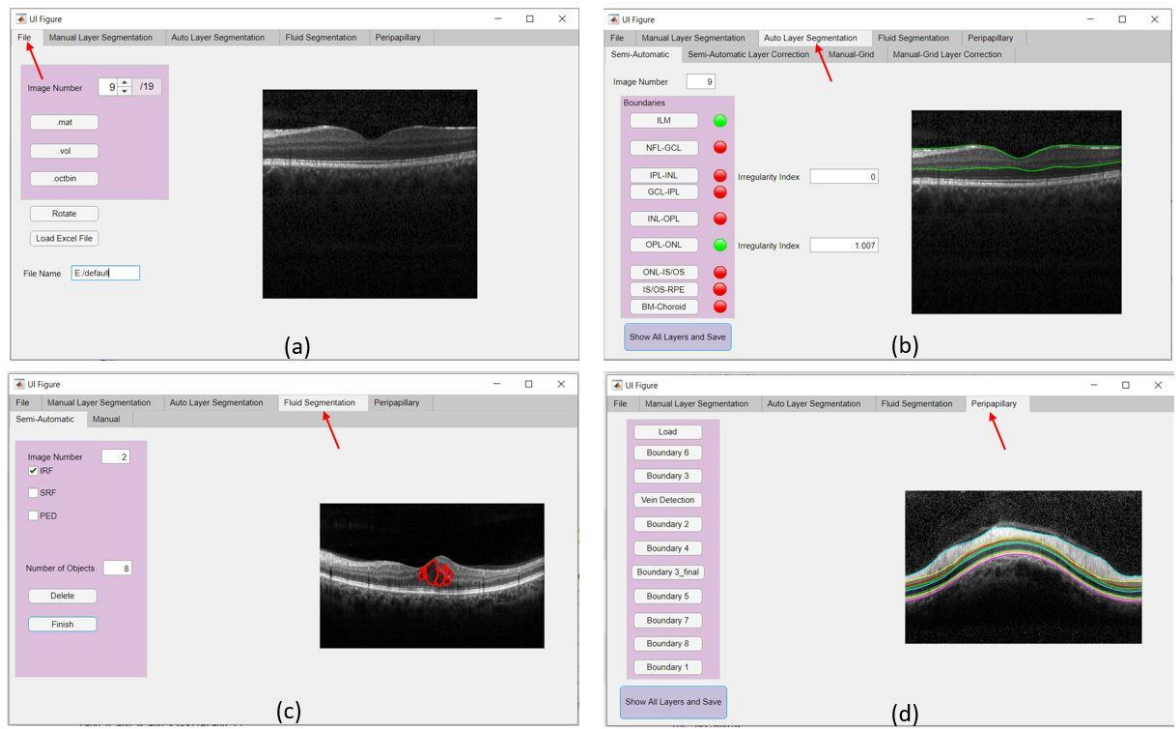

Figure 1. The proposed software's major tabs   (a) The File tab     (b) The Auto Layer Segmentation tab
(c) The Fluid Segmentation tab     (d) The Peripapillary tab

### The devised Livelayer algorithm

Our suggested algorithm is a graph-based semi-automatic method, called "Livelayer", designed for straightforward segmentation of retinal layers and fluids. This method is chiefly based on Dijkstra's Shortest Path First (SPF) algorithm [28] and the Live-wire function [29], together with some preprocessing operations on the to-be-segmented images. The Dijkstra's algorithm and its interactive variant, the live-wire, find the shortest path between a pair of nodes (pixels) in a graph (that is constructed from the original image). Application of the conventional Live-wire over the original B-Scan is not capable of following the OCT boundaries due to their weak and vague appearance and it is essential to isolate and strengthen each individual boundary before being fed into the live-wire algorithm. The proposed Livelayer is created on the basis of the original live-wire and is applied to different processed versions of the original B-Scan which have the ability to sharpen the desired boundary. We detect these best sharpened boundaries by utilizing diverse methods of edge detection and morphological operations [30-34]. Livelayer can also be applied over B-Scans containing fluid objects and circumpapillary scans. Fig. 3 goes into details about the pre-processing block exploited in the Livelayer. As can be seen, depending on each boundary's brightness and location in the B-Scan, an apt background image for applying the Livelayer is designed using different image processing techniques. Additionally, the background image for fluid segmentation is acquired through implementation of an edge detection function followed by a morphological operation.

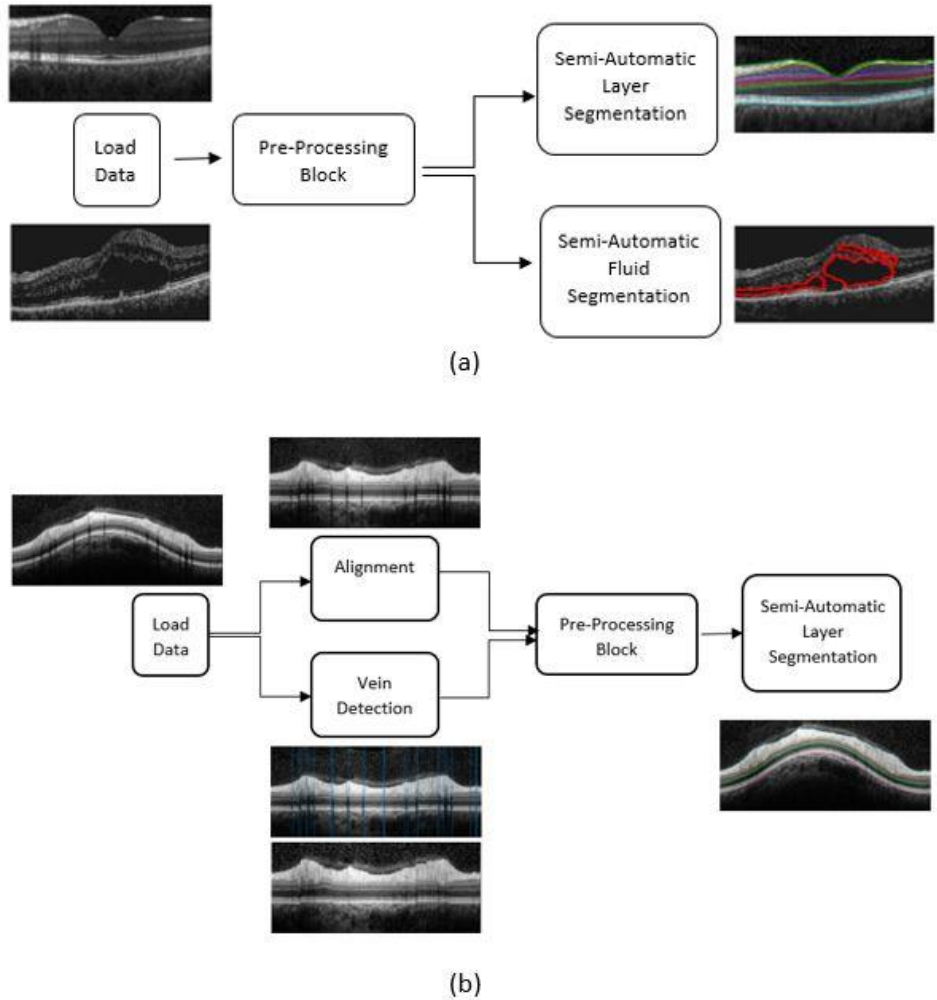

Figure 2. Livelayer software's scheme (a) Macular layer and fluid segmentation sections - After loading a proper data set, the software's pre-processing block finds an appropriate background image on which the Livelayer could be applied and then, outputs the segmented image. (b) Peripapillary layer segmentation section - After loading a proper peripapillary data, the software aligns the B-Scan and then, tries to omit its veins as much as possible. An appropriate background image for applying the Livelayer is detected and the corresponding segmented B-Scan is outputted.

**Validation of the proposed software**

Our dataset in this study generally consists of the normal OCT B-Scans acquired from healthy controls and abnormal OCT B-Scans that are related to DME patients. The former group of data is used for validation of our suggested method's layer segmentation section and the latter is for the algorithm's fluid segmentation part. In the following paragraphs, we go into concise explanations for the procedure adopted to assess our algorithm's performance and how it can be utilized as a useful tool for clinical purposes.

**Layer Segmentation Section**

The retinal boundaries in macular OCT scans that our software is able to segment include: inner limiting membrane (ILM), boundary between retinal nerve fiber layer (RNFL) and ganglion cell layer (GCL), boundary between GCL and inner plexiform layer (IPL), boundary between IPL and inner nuclear layer (INL), boundary between INL and outer plexiform layer (OPL), boundary between OPL and outer nuclear layer (ONL), boundary between ONL and photoreceptor layer, boundary between photoreceptor layer and retinal pigmented epithelium (RPE) and finally, outer level of RPE. The peripapillary section could segment all these boundaries except for the GCL-IPL.

In order to perform a validation of our suggested method, two ophthalmologists and an engineer familiar with the concept of OCT images delineated the three most clinically important retinal boundaries (ILM, IPL-INL, OPL-ONL) on the healthy controls' dataset and with the aid of two discrete methods. In the first stage, we trained these three independent individuals to grid-manually segment the destined retinal boundaries, which is considered to be an impotent segmentation procedure as opposed to the Livelayer algorithm. Following that, three graders were also asked to segment the same B-Scans with the Livelayer. The main manual method in this software which was created utilizing the freehand function was intolerably time-consuming and susceptible to noticeable errors, which is why it was substituted with our grid-based algorithm.

To properly quantify the agreement and repeatability of the proposed semi-automatic method, we need to first define a basis for comparison that is the unsigned boundary errors. The unsigned boundary error for retinal boundaries is calculated by:

$$uerror = \frac{\sum_{i=1}^{w} abs(x_i - y_i)}{w} \tag{1}$$

where $w$ denotes the B-Scans's width in pixels and $x_i, y_i$ represent the $i^{th}$ point of the two acquired boundaries. The lower the value of the unsigned boundary error is, the more proximity we observe between the two boundaries.

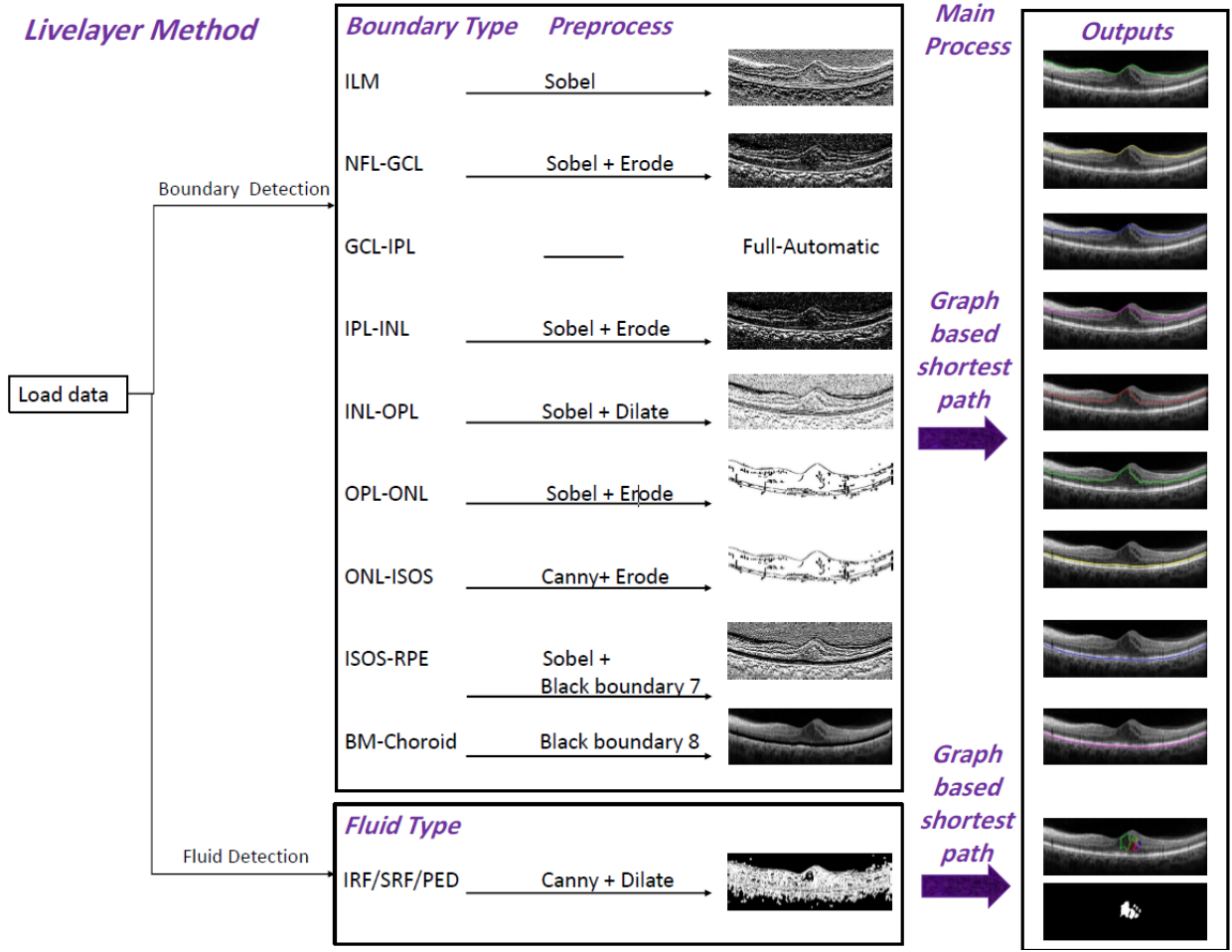

Figure 3. An overview of the Livelayer's pre-processing block – Various operations conducted on each boundary depending on their brightness and location in the B-Scan

Therefore, we found the average location of each boundary grid-manually outlined by three graders on each OCT B-Scan, and assumed it to be the ground truth in our studies. Next, we calculated the unsigned error between the ground truth and each of our three grid-manual delineations to achieve the inter-rater variability for this method. This ground truth data was then compared to each of our three semi-automatic data for obtaining both the inter-rater variability of our algorithm and determine whether Livelayer could be substituted with the manual segmentation approach. To quantify our measurements, the Bland-Altman plots for each boundary were sketched, obtaining the level of agreement between each grader's semi-automatic segmentation and the gold standard data. For determining the repeatability of the Livewire, one of our observers segmented the 50 normal B-Scans for the second time a week later and the unsigned errors between these two measurements were computed.

To indicate the timesaving capability of the software, the required time for segmenting each boundary was measured by utilizing both the semi-automatic Livelayer and the grid-manual methods. Additionally, the mean number of needed clicks by the user in Livelayer method were figured out. It should be noted that the number of clicks in Livelayer depends both on the image quality and the intrinsic characteristics of that boundary on the image.

This section also assigns a particular field for computation of the irregularity index for IPL and OPL within the semi-automatic division. This index is set for quantitative smoothness evaluation of a specific layer in macular OCT images and is defined according to this equation:

$$Irregularity\ Index = \frac{L_r}{L_b} \tag{2}$$

where $L_r$ is the Euclidean distance between the beginning and the end of that boundary and $L_b$ is the length of the intended detached boundary and

**Fluid Segmentation Section**

This method is not primarily designed to compete with other full-automatic fluid segmentation available methods, but to be served as a handy tool for production of an outlined dataset which could be later required for training the automatic algorithms. For obtaining the agreement and repeatability of this division, two independent persons, an ophthalmologist and a senior ophthalmology resident were trained to both manually and semi-automatically localize the intraretinal and subretinal fluid objects on the DME dataset.

Here, we supposed the dice coefficient to be a proper basis for comparison of the located fluid objects. For retinal fluids, the dice coefficient [35] which gauges the basic similarity between two mask images in which the fluid objects' regions of interest (ROIs) are apparent is calculated by:

$$dice = \frac{2\ (X \cap Y)}{(X \cup Y)} \tag{3}$$

where $X$ and $Y$ are the two intended mask images. The smaller values of dice coefficient indicate less similarity between two set of identified fluids.

This coefficient was computed for the two observers' manually segmented fluids as well as for one observer's manual and semi-automatic results to resemble inter-observer errors and the method's performance. Because the two graders did not agree on the overall number of fluids on each B-Scan, a particular threshold was defined and marginally small fluid objects were eliminated according to that threshold. For evaluating the repeatability of this section, one of the examiners delineated the fluid objects on the DME dataset for another time 6 weeks later and the dice coefficients for these assessments were measured.

## Results

### Efficiency in time and the required number of clicks

For assessment of our recommended method's efficiency in the matter of time and the needed number of clicks, all the 50 normal macular B-Scans that were segmented semi-automatically and twenty of grid-manually segmented B-Scans were involved. Table 2 demonstrates a precise comparison between the semi-automatic and the grid-manual segmentation methods' acquisition time and depicts the average number of clicks that the Livelayer required. The grid-manual segmentation in this part was conducted by splitting each B-Scan into 10 portions and using the layer correction functionality of the software around the fovea. Hence, the grid-manual algorithm needed at least 10 clicks in the best case.

Table 2. A comparison between the required amount of time (in seconds) and the number of clicks for Livelayer and grid-manual

| Boundary | Time (Livelayer) | Time (Grid-manual) | Number of Clicks (Livelayer) |
|---|---|---|---|
| ILM | 5 | 27.69 | 4.5 |
| IPL-INL | 10.86 | 22.57 | 5.9 |
| OPL-ONL | 15.57 | 22.12 | 8.4 |

### Feasibility on the DME dataset

As we mentioned earlier, Livelayer also performs pretty well on the B-Scans obtained from patients with DME, and despite the existence of large cysts in these images, it could segment relatively disorganized retinal layers generally caused by retinal disorders such as Diabetic Retinopathy. Fig. 4 presents three B-Scan samples segmented by our introduced algorithm. This issue proves the feasibility of our software to be employed on specific kinds of datasets that could not be automatically segmented by the Heidelberg system.

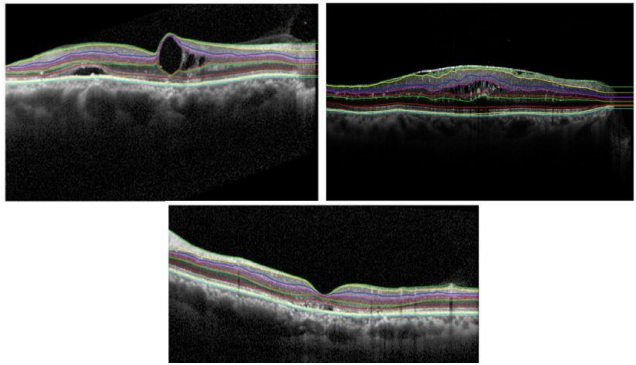

Figure 4. Three different macular B-Scans obtained from patients with DME and segmented by the Livelayer algorithm

**Agreement and repeatability analyses**

Analysis of agreement and repeatability for the Livelayer was performed using 50 normal macular B-Scans. In Fig. 5 the three boundaries are marked by our observers in both modes. In Table 3, the unsigned errors between the gold standard segmentation and each examiner's grid-manually labeled data, the gold standard and each examiner's semi-automatically labeled data and the two independent semi-automatic segmentation data for one grader is represented.

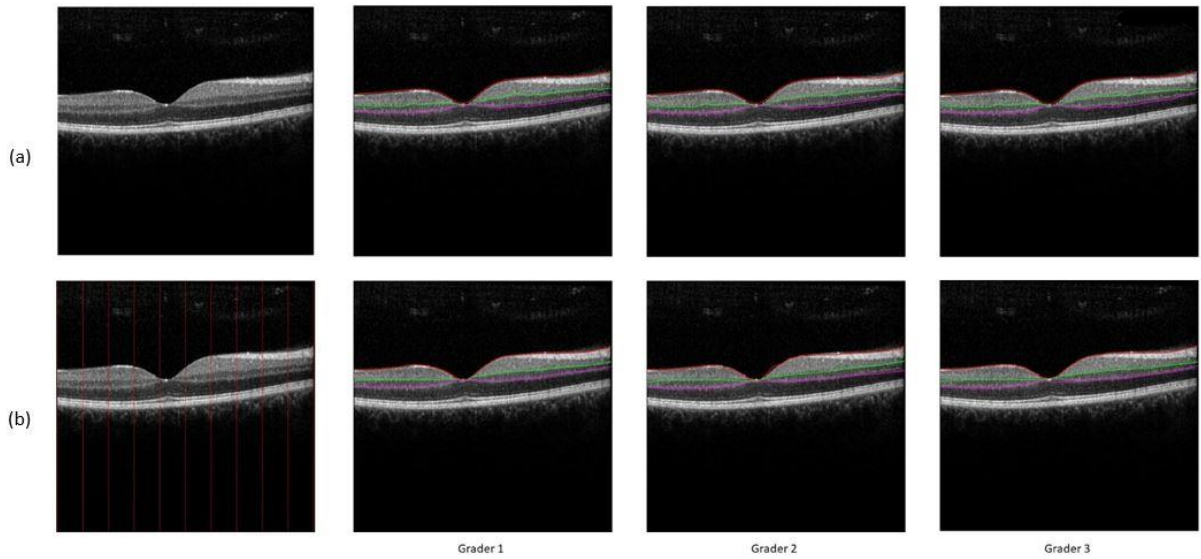

Figure 5. A representation of three segmented boundaries by each grader in the (a) Semi-automatic mode (b) Grid-manual mode

Table 3. Unsigned errors (in pixels) between the gold standard and each observer's grid-manual and semi-automatic segmentation as well as the intra-observer errors

|  | ILM | IPL-INL | OPL-ONL |
|---|---|---|---|
| Gold-Observer1(Grid) | 0.67 | 1.29 | 2.38 |
| Gold-Observer2(Grid) | 0.74 | 1.02 | 1.82 |
| Gold-Observer3(Grid) | 0.8 | 0.88 | 1.03 |
| Gold-Observer1(Semi) | 2.75 | 1.86 | 1.85 |
| Gold-Observer2(Semi) | 2.7 | 1.9 | 2.24 |
| Gold-Observer3(Semi) | 2.65 | 1.94 | 2.11 |
| Intra-observer (Repeatability) | 0.37 | 0.74 | 1.02 |

Bland-Altman plots for each pair of the semi-automatic segmentation coming in Table 3 is drawn in Fig. 6. As can be seen, delineation of the ILM showed 96% total agreement for graders 1, 2 and 94% for grader 3. The total agreement between the Livelayer and the gold standard data for IPL-INL was also acceptable, being 98% for grader 1 and 96% for graders 2,3. Among these three boundaries, the least amount of agreement was expected for OPL-ONL because of the existence of some peculiar structures that frequently occur under this boundary, called the Henle's fiber. However, despite the occurrence of these irregularities, a good agreement between two methods could be observed for this boundary which was comparable to the other two, 94% for graders 1,3 and 92% for grader 2. In general, the Bland-Altman plots demonstrated no significant discrepancy between the two segmentation methods and our suggested semi-automatic algorithm could be reliably substituted with typical manual segmentation methods.

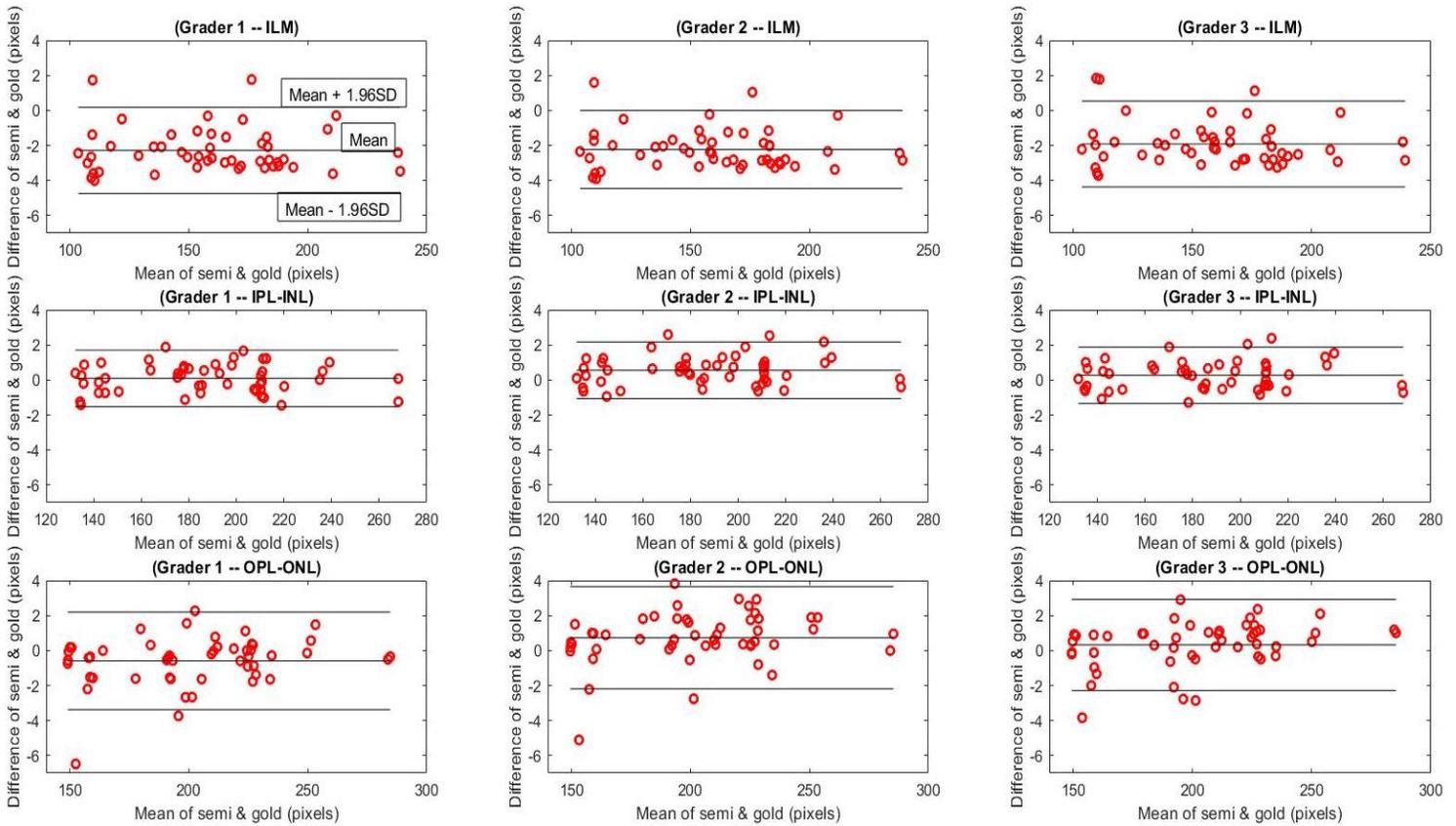

Figure 6. The Bland-Altman plots for correlating our proposed semi-automatic algorithm with the gold standard data (average of grid-manually delineated B-Scans by three examiners) – Each row presents the B&A plots for each specific boundary and each column is assigned to the boundaries segmented by each examiner.

The software's fluid segmentation performance was evaluated on 20 B-Scans of DME patients. Our proposed algorithm was proven to surpass the manual segmentation method (freehand function) by obtaining dice scores of 0.854 and 0.743 for one grader's manual and semi-automatic fluid (IRF and SRF) localization and two graders' manual localization, respectively. Moreover, the fluid objects that were delineated by one of our graders for two times in 6 weeks showed 89.3% coincidence according to the calculated dice scores. So, the proposed semi-automatic software worked with good precision on the fluid objects concerning its inter-rater variability and repeatability as well.

## Conclusion and Discussion

In this paper, we introduced a semi-automatic segmentation software tool which can be used for segmentation of layers and fluids in OCTs of both normal participants and patients with DME. Then, as well as introducing a powerful clinical application of this program which is the calculation of the irregularity index of a specific boundary, we evaluated three of the software's performance qualities including its segmentation agreement, repeatability and efficiency (in terms of time and complexity). With the purpose of achieving a reliable estimation of the software's layer segmentation agreement, three independent individuals segmented the layers in both grid-manual and semi-automatic modes. The agreement was finally found by defining the gold standard data (which was the average of grid-manual measurements among three graders) and then comparing this data with the measured values pertinent to the grid-manual segmentation and the same type of values for the semi-automatic results. Our measurements for showing the total agreement between the grid-manual and gold standard data were carried out utilizing the unsigned errors. Those measurements for showing the total agreement between the gold standard data and Livelayer were quantified using both the unsigned errors and the Bland-Altman plots. In the next stage, the inter-rater variability of the algorithm's fluid segmentation section was computed with the aid of dice coefficients. Using this value, we figured out how much the mask images that were manually and semi-automatically segmented by one individual overlapped with each other, and compared their relevant dice coefficient with the one obtained from semi-automatically detected fluids by two graders. In addition to that, we asked one observer to segment the layers and fluid objects twice and in a determined time interval to attain the repeatability of both sections, assuming that other conditions remained constant during this study.

Examination of the software's efficiency, conducted on 50 normal macular B-Scans, indicated that the proposed Livelayer algorithm, effectively integrated into the Livelayer software, was way less complicated than a common grid-manual segmentation method and thus, required a short period of time for delineation of layers in macular OCT images. Furthermore, the proposed algorithm's performance level was evaluated by applying the Livelayer on the former set of B-Scans, deducing that even though the algorithm might vary slightly in its performance quality (i.e. the required number of clicks) depending on the image's innate qualities, it worked more efficiently than a typical grid-manual method and almost always needed much less number of clicks. Testing the algorithm over the OCTs related to patients with the diagnosis of DME showed its practicality to segment unclear and detached retinal layers with a nearly high precision in comparison to the segmented data collected from the Heidelberg Eye Explorer (HEYEX) system.

Segmentation of retinal OCT images is an essential technique used in a variety of applications ranging from clinical to research studies. Due to the inherently layered structure of retina and changes in these layers in the presence of mild to severe retinal diseases such as diabetic retinopathy, studying these variations will help to understand the pathogenesis of various retinal disorders. Morphological analysis of each of these layers has yet to be comprehensively done except in a limited number of studies because of the large amount of time it takes to segment each layer manually. Therefore, designing a fast and reliable method that could overcome this obstacle should be of considerable importance. Manual segmentation methods are indeed tedious and complex, and full-automatic segmentation is actually deficient in its adaptability to human error correction. Semi-automatic algorithms are, however, capable of being the most preferred kind of ocular image segmentation methods among researchers and clinicians.

In Table 4 a set of elected automatic and semi-automatic segmentation methods on retinal OCT data are compared with our suggested method in terms of "simplicity and availability", "processing time", "capability in handling errors" and "capability in handling huge amount of data". Lee et al suggested a full-automatic software package, OCT Explorer (the Iowa Reference Algorithm) [36-38] for segmentation of 12 retinal boundaries which is 3 more boundaries than what our software is capable of identifying. In contrast to our software's error correction functionality that works independently from other segmented boundaries, in OCT Explorer, refining a single boundary highly affects other segmented boundaries on both the corresponding B-Scan and the other ones existing in the whole dataset. Furthermore, OCT Explorer fails to precisely find all boundaries of cystic B-Scans due to their obscure appearance on the image whereas the Livelayer's ability to segment different types of OCTs was argued in the previous sections of this study. OCTMarker [20] is an integrated software package that could only detect up to 3 retinal boundaries automatically. EdgeSelect [16] and SAMIRIX [20] are two semi-automatic methods for OCTA segmentation which are not available for free and their possible function is speculated as reported by their papers accordingly. Unlike the EdgeSelect that only segments 4 retinal boundaries, SAMIRIX is capable of segmenting 9 retinal boundaries, but accepting just one file format (.vol). However, Livelayer software reads 3 distinct input file formats making the user feel no need for volume converter software tools.

Table 4. A comparison between previously suggested segmentation algorithms and the Livelayer software

| Algorithm's Name | Simplicity and Availability | Processing Time | Handling Errors | Handling Huge Data | Extra Notes |
|---|---|---|---|---|---|
| OCT Explorer (Iowa Reference Algorithm) [36-38] | Almost simple Freely available | Automatic (less than 10 seconds) | Including an error correction functionality - Editing a boundary on one slice considerably affects other boundaries and slices. | Yes | An integrated software package- Accepting various file formats - Not very good at low-quality, vein containing OCTs |
| OCTMarker [20] | Almost simple Freely available | Automatic (less than 5 seconds) | Including an error correction functionality | Yes | An integrated software package– Capable of segmenting very few retinal layers (approximately 2 or 3 layers) |
| EdgeSelect [16] | Not available | Semi-automatic Not discussed | - | - | Not an integrated software package– Only 4 boundaries segmented |
| SAMIRIX [20] | Not available | Semi-automatic Not discussed | - | - | An integrated software package– Accepting only one file format – Capable of segmenting 9 retinal boundaries |
| **Livelayer (our proposed)** | **Quite simple Freely available** | **Semi-automatic (roughly 60 seconds)** | **Including an error correction functionality** | **Yes** | **An integrated software package– Accepting 3 file formats -Capable of segmenting 9 retinal boundaries – Exact in low-quality OCTs** |

The most remarkable superiority of our software over other previously presented methods is its feasibility to firstly, accommodate different methods of layer and fluid segmentation required by someone who may not be expert in the concepts of image processing like an ophthalmologist or a neurologist and secondly, generate the output in either a .jpg format (image file), or a .xlsx one (excel files) which could be specially advantageous while working with bulk data. Another major capability of the Livelayer software is the different filtering processes adaptable to each layer's properties it employs. Other proposed algorithms generally apply a particular set of filtering on the to-be-segmented images which would always be constantly used regardless of the resolution of the input data as well as the quality of retinal layers. The Livelayer, on the other hand, considers each of these factors for its filtering stage by applying independent filters on each boundary and including an editable file in which the filtering parameters could be adjusted. Fig. 7 displays a hierarchy of how the software creates the corresponding folders for each section and how the software's output information is saved. In addition to that, whereas the Heidelberg system from which our peripapillary data is acquired, is not capable of segmenting all the peripapillary layers except for the pRNFL, our software could accurately segment 8 peripapillary boundaries after omitting their veins and applying a set of pre-processing operations on them. The HEYEX system is not also able to exactly pinpoint the detached retinal layers occurring on the OCT B-Scans related to DME or other akin ailments, yet the Livelayer could tackle this problem with a few more than average click numbers. On account of being an open-source product, it can be upgraded to newer versions by either adding more essential tabs or improving the current ones to achieve a specific performance quality like a better denoising or to be tailored for diverse types of OCTs (e.g. rat retinal OCTs), and thus, to become even more efficient.

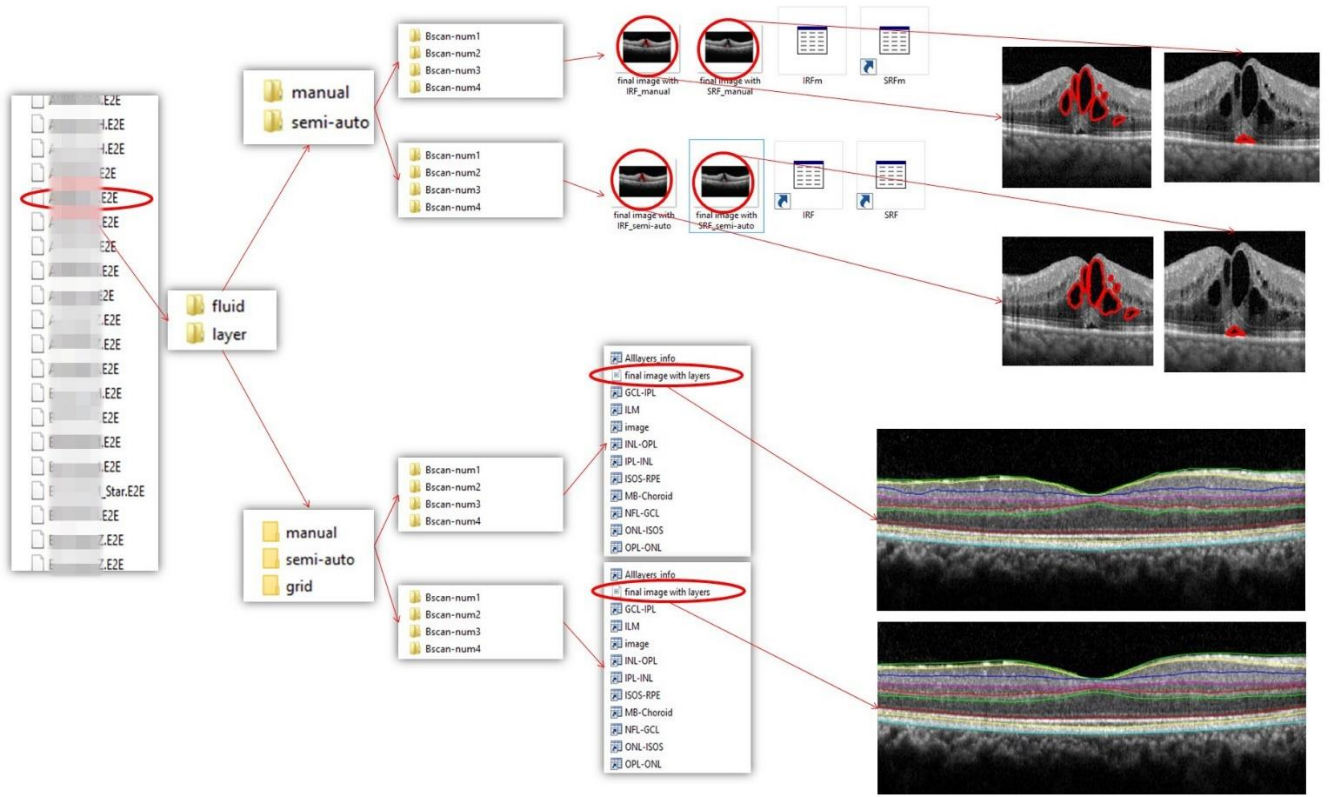

Figure 7. The structure of the DME dataset we worked on and a hierarchical model for the discretely generated folders constituting the software's output

As it is evident from the results section, the proposed software is reasonably efficient in being time-effective and straightforward. The semi-automatic segmentation time and the number of required clicks for each of the studied retinal boundaries was far less than what they were for the manual-grid method. Regarding this analysis, ILM was the most rapidly- and easily-segmented layer with the biggest disparity between its corresponding values for two intended methods, and OPL-ONL needed considerably longer time and a greater number of clicks to be segmented. This is because of the irregular Henle's fiber appearing under the OPL, especially its foveal portion causing the Livelayer not to rightly discriminate between this layer and the ONL. The software also showed a good level of agreement between each of the manual-grid and semi-automatic methods, and the gold standard data. Comparing the gold standard data with each grader's grid-manual delineation yielded less than 1 pixel unsigned errors for ILM, gradually increasing to reach near 2 pixel errors for OPL-ONL. On the other hand, since our proposed method locates ILM slightly higher than its actual location, the unsigned errors between the gold standard data and the semi-automatic segmentation of ILM was larger than that of other boundaries. This error for other two boundaries fluctuated around 2 pixels with generally smaller values for IPL-INL. Analysis of repeatability which is also shown in Table 3 achieved pretty small errors for ILM and IPL-INL, but a near 1 pixel error for OPL-ONL. The Bland-Altman plots which are brought to quantify the level of agreement achieved total agreements for all boundaries with quite analogous values for ILM and IPL-INL and with an approximately 2-4 per cent decline in these values for OPL-ONL. Concerning the fluid segmentation section, the dice score coefficients for both inter-rater variability and repeatability studies were at an acceptable level and proved that this algorithm could be used as a substitute for manual methods.

## Author Contributions

Below are the authors of this manuscript along with their contributions to this study.

| | MM | ZS | EK | HRE | TM | HR | HM | AD | MA | RK |
|---|---|---|---|---|---|---|---|---|---|---|
| Conceptualization | Major | | | | | Major | | | | Major |
| Methodology | Major | | Supporting | | | | | | | Major |
| Software Design | Major | Major | | | | | | | | |
| Validation | Major | | | | | | | | | Major |
| Formal Analysis | Major | Major | | | Major | | | | | |
| Investigation | Major | | | | | | | | | Major |
| Data Segmentation | | | Major | Major | | | Supporting | | | |
| Labeled Data Checking | | | Major | Major | | | | Supporting | Supporting | |
| Visualization | Major | Major | | | | | | | | |
| Writing - Original Draft | Major | | | | | | | | | |
| Writing - Review and Editing | | | Major | Major | | | Supporting | | | Major |
| Supervision | | | | | | | | | | Major |
| Levels of contribution : Major, Supporting | | | | | | | | | | |

## Competing interests

The authors declare no competing interests.

## Figure Legends

Figure 1. The proposed software's major tabs   (a) The File tab    (b) The Auto Layer Segmentation tab   (c) The Fluid Segmentation tab    (d) The Peripapillary tab

Figure 2. Livelayer software's scheme   (a) Macular layer and fluid segmentation sections - After loading a proper data set, the software's pre-processing block finds an appropriate

background image on which the Livelayer could be applied and then, outputs the segmented image. (b) Peripapillary layer segmentation section - After loading a proper peripapillary data, the software aligns the B-Scan and then, tries to omit its veins as much as possible. An appropriate background image for applying the Livelayer is detected and the corresponding segmented B-Scan is outputted.

Figure 3. An overview of the Livelayer's pre-processing block – Various operations conducted on each boundary depending on their brightness and location in the B-Scan

Figure 4. Three different macular B-Scans obtained from patients with DME and segmented by the Livelayer algorithm

Figure 5. A representation of three segmented boundaries by each grader in (a) Semi-automatic mode   (b) Grid-manual mode

Figure 6. The Bland-Altman plots for correlating our proposed semi-automatic algorithm with the gold standard data (average of grid-manually delineated B-Scans by three examiners) – Each row presents the B&A plots for each specific boundary and each column is assigned to the boundaries segmented by each examiner.

Figure 7. The structure of the DME dataset we worked on and a hierarchical model for the discretely generated folders constituting the software's output